\documentclass[12pt]{iopart}

\usepackage[english]{babel}
\usepackage{mathrsfs}
\expandafter\let\csname equation*\endcsname\relax
\expandafter\let\csname endequation*\endcsname\relax
\usepackage{amsmath}
\usepackage{amssymb}
\usepackage{amsthm}
\usepackage{amsfonts}
\usepackage{BOONDOX-cal}
\usepackage{graphicx}
\usepackage{braket}
\usepackage{subcaption}
\usepackage{multirow}
\usepackage[table]{xcolor}
\usepackage{bbm}
\usepackage{perpage}
\usepackage{colortbl}
\usepackage{graphicx}
\usepackage{subcaption}
\usepackage{tikz}
\usetikzlibrary{calc}
\usepackage{cite}

\definecolor{nicepurple}{HTML}{97a0cf}
\definecolor{lightgreen}{HTML}{88BFB7} 
\definecolor{lightergreen}{HTML}{88BFB7} 
\definecolor{cgreen}{HTML}{88BFB7}
\definecolor{melon}{HTML}{FCBCB8}
\definecolor{puce}{HTML}{C08497}
\definecolor{palegreen}{HTML}{A9DAD2}

\renewcommand{\newline}{\\\\\noindent}

\newcommand{\jmax}{m_{max}}

\DeclareMathOperator{\sign}{sign}

\newcommand{\dataset}[2]{\left\lbrace \mathbf{#1}, #2\right\rbrace}
\newcommand{\classifier}{\mathcal{C}}
\newcommand{\regressor}{\mathcal{R}}
\newcommand{\meta}{\mathcal{P}}
\newcommand{\newparagraph}{\\\\\noindent}
\renewcommand{\sign}{\mathrm{sgn}}

\newcommand{\code}[1]{\texttt{#1}}
\newcommand{\logtenj}[1]{\mathrm{log10j(#1)}}
\newcommand{\tenjsymbol}{\left\lbrace 10j \right\rbrace}

\begin{document}

\title[]{Deep learning spinfoam vertex amplitudes: the Euclidean Barrett-Crane model}

\author{Hanno Sahlmann $^1$, Waleed Sherif $^2$\footnote{Author to whom any correspondence should be addressed}}

\address{Institute for Quantum Gravity, Department of Physics, Friedrich-Alexander-Universit\"{a}t Erlangen-N\"{u}rnberg (FAU), Staudtstraße 7, 91058 Erlangen, Germany}

\ead{$^1$ hanno.sahlmann@fau.de, $^2$ waleed.sherif@fau.de}

\vspace{10pt}
\begin{indented}
\item[] \today
\end{indented}
\begin{abstract}
Spinfoam theories propose a well-defined path-integral formulation for quantum gravity and are hoped to provide the dynamics of loop quantum gravity. However, it is computationally hard to calculate spinfoam amplitudes. The well-studied Euclidean Barrett-Crane model provides an excellent setting for testing analytical and numerical tools to probe spinfoam models. We explore a data-driven approach to accelerating spinfoam computations by showing that the vertex amplitude is an object that can be learned from data using deep learning. We divide the learning process into a classification and a regression task: Two networks are independently engineered to decide whether the amplitude is zero or not and to predict the precise numerical value, respectively. The trained networks are tested with several accuracy measures. The classifier in particular demonstrates robust generalisation far outside the training domain, while the regressor demonstrates high predictive accuracy in the domain it is trained on. We discuss limitations, possible improvements, and implications for future work. 
\end{abstract}

\section{Introduction}

Loop quantum gravity (LQG) is a canonical quantisation program for general relativity (GR) that attempts to keep general covariance as manifest as possible \cite{Rovelli:1997yv,Thiemann:2001gmi,Ashtekar:2004eh,Han:2005km}. In its kinematical sector, the theory is well-formulated: a Hilbert space $\mathcal{H}_{kin}$ spanned by spin-network states - labelled by graphs whose oriented edges carry irreducible SU(2) representations and whose vertices carry SU(2) invariant intertwiners \cite{Ashtekar:2004eh}. The dynamics, however, is far more elusive. It is encoded in constraints - operators imposing equations (or, in the Master constraint approach \cite{Thiemann:2003zv,Dittrich:2004bn,Thiemann:2005zg} a single equation) on the physical states. Solutions are timeless a priori. Finding solutions, and giving them a spacetime-interpretation remains a formidable open problem despite various efforts (see for example \cite{Thiemann:1996aw,Thiemann:1996av,Varadarajan:2022dgg,Bojowald:2001xe,Ashtekar:2006wn,Guedes:2024zbu,Guedes:2024duc,Sahlmann:2024pba,Sahlmann:2024kat}).
\newparagraph
Path integral formulations of gravity have been considered for a long time, since they avoid the space-time split inherent in the canonical approach. Spinfoam (SF) theories (for example, see \cite{Reisenberger:1994aw,Reisenberger:1996pu,Rovelli:2014ssa,Perez:2012wv,Baez:1997zt,Bambi:2023jiz,Engle:2023qsu,Livine:2024hhc,Kaminski:2009fm,Bahr:2010bs}) in particular originated from the observation that gravity can be obtained by breaking the symmetries of a topological gauge theory. These theories side-step part of the difficulties of canonical LQG by constructing transition amplitudes between spin-network states as a sum over (discrete quantum) histories. 
\newparagraph
To regularise the gravitational path integral, one triangulates the manifold by a simplicial complex $T$ and works on its dual 2-complex $\Delta^*$. A spinfoam is precisely such a 2-complex whose faces $f$ are labeled by irreps $\pi_f$ of the gauge group and whose edges $e$ carry intertwiners $i_e$ between the irreps of the faces meeting in $e$. The regularised partition function can be written in factorised form \cite{Rovelli:2014ssa,Perez:2012wv}
\begin{equation}
    Z_{\Delta^*} = \sum_{\pi_f, i_e} \prod_f A_f(\pi_f) \prod_e A_e(i_e) \prod_v A_v(\pi_f, i_e),
\end{equation}
where $v$, $e$ and $f$ are dual to 4-simplices, tetrahedra and triangles of $T$ respectively. Mirroring the role of a vertex in ordinary Feynman diagrams, the vertex amplitude $A_v$ encodes the local dynamics of quantum geometry \cite{Engle:2007uq}. The precise nature of the gauge group, $\pi_f$, $i_e$ and $A_v$ depends on the spacetime dimension, signature, and theory. 
\newparagraph
A serious problem is that $Z_{\Delta^*}$ is very hard to calculate in practice. The sum contains infinitely many terms in principle. Moreover, even calculating the vertex amplitude $A_v$ can be difficult, since it typically is a highly oscillatory function which involves multidimensional sums or integrals in its definition and depends on the data of the vertex-spin network, i.e., the $\pi_f$- and $i_e$-label surrounding $v$.
\newparagraph
The development of various numerical methods \cite{Dona:2022yyn}, each tailored to suit different regimes, have played a pivotal role in driving advancements in spinfoam models. Highly optimised, high-performance computing libraries such as \texttt{sl2cfoam} \cite{Dona:2019dkf} and its successor \texttt{sl2cfoam-next} \cite{Gozzini:2021kbt,Dona:2022dxs} enable efficient evaluation of Engle-Pereira-Rovelli-Livine (EPRL) amplitudes \cite{Engle:2007uq,Engle:2007wy} and have been used to study various aspects of spinfoam models \cite{Dona:2020tvv,Dona:2022vyh,Frisoni:2022urv}. Monte Carlo methods, enhanced by Lefschetz thimbles techniques to deform integration contours, have also been utilised in the context of spinfoams \cite{Han:2020npv}. Different sampling methods over representation labels, such as importance sampling and random sampling of the bulk spins as well as utilising generative flow networks, have been utilised to further enhance convergence of spinfoam amplitudes involving several simplices or efficiently computing expectation values of observables \cite{Dona:2023myv,Bunao:2024qwm}. Monte Carlo methods were also used to compute the vertex amplitude for a given set of coherent states as boundary data \cite{Steinhaus:2024qov}. Recently, tensor-network methods and techniques from many-body quantum physics have been utilised to significantly reduce both the computational complexity and memory requirements for computing vertex amplitudes for both SU(2) and EPRL spinfoam vertex amplitudes \cite{Asante:2024eft}.
\newparagraph
On the large-spin or asymptotic regime, numerical programs such as the complex critical points program \cite{Han:2021kll,Han:2023cen,Han:2024lti} have been developed to identify semi-classical geometries in the limit of large representations while a hybrid program was proposed in \cite{Asante:2022lnp} to bridge between the quantum and the semi-classical regimes. Numerical methods \cite{Bahr:2015gxa,Bahr:2016hwc}, based on symmetry reductions, have also been employed to the study of the renormalisation aspects of spinfoam models. 
\newparagraph
Despite this considerable effort in both analytical and numerical approaches (see also \cite{Asante:2020qpa,Asante:2021zzh,Asante:2020iwm,Dittrich:2021kzs,Dittrich:2022yoo,Borissova:2022clg}), the computation of vertex amplitudes remains a challenging problem. This motivates the exploration of alternative approaches to address the inherent computational complexity. The goal of the present work is to establish a new approach to accelerating spinfoam computations by showing that the vertex amplitude is an object that can be learned from data using deep learning. In the simplest form, a trained neural network would approximate $A_v$ by interpolating from training data. In a more ambitious form, the neural network would learn to extrapolate much beyond its training domain in certain spin regimes. Further, another aim of this work is to complement the numerical implementations of exact analytical methods by helping identifying dominant configurations, guiding importance sampling and enabling efficient pre-selection in possibly large parameter spaces.
\newparagraph
While neural networks have been utilised in the context of spinfoam computations to calculate expectation values of observables \cite{Bunao:2024qwm} by learning the boundary configurations which contribute large amplitudes, the current work, as far as we are aware, is the first of its kind where the vertex amplitude itself is learned using a neural network. Consequently, we aim for a proof-of-principle, not an application to state-of-the-art spinfoam models. To keep things as non-technical as possible, we consider a very well studied and non-trivial model, the Euclidean Barrett-Crane-model (BC-model) \cite{Barrett:1997gw,Barrett:1999qw,Baez:1999sr,Reisenberger:1998bn,Livine:2001jt}. While interesting, this model is nowadays understood to be unphysical \cite{Alesci:2007tx}. 
The action for the path integral for this theory derives from the Plebanski (or $BF$-theory plus simplicity constraints) action \cite{Plebanski:1977zz}
\begin{equation}
    S_{Plebanski}[B, \omega, \lambda] = \int \tr \left[B \wedge F(\omega) + \lambda B \wedge B\right],
\end{equation}
where $B$ is an so(4)-valued 2-form, $\omega$ is an SO(4) connection and $\lambda$ a matrix-valued Lagrange multiplier enforcing the simplicity constraints that reduce topological $BF$-theory to the familiar Palatini action \footnote{For Lorentzian signature one replaces SO(4) with SO(1, 3) and must deal with non-compact representations. We confine ourselves to the Euclidean sector throughout this work.}\cite{Baez:1999sr,Livine:2001jt}. 
\newparagraph
Exploiting $\mathrm{Spin}(4) \cong \mathrm{SU}(2)_+ \times \mathrm{SU}(2)_-$, the BC-model strongly imposes the simplicity constraints, thereby restricting every face to a balanced representation $(j_+, j_-)=(j, j)$. While this drastic reduction yields a manageable state sum, it also freezes the intertwiner degrees of freedom, eliminating physical degrees of freedom and effectively rendering the model unphysical \cite{Alesci:2007tx}. More refined models such as the EPRL or Engle-Pereira-Rovelli-Livine-Freidel-Krasnov (EPRL-FK) \cite{Freidel:2007py} models impose the constraints weakly. They reintroduce an SU(2) intertwiner label, depend non-trivially on the Barbero-Immirzi parameter $\gamma$ and reproduce the BC-model (intertwiner) in the limits $\gamma \rightarrow \pm \infty$ \cite{Kaminski:2009fm}.
\newparagraph
We will consider the BC-model on a simplicial complex.  It's vertex amplitude then  takes the characteristic and compact form \cite{Christensen:2009bi}
\begin{equation}
\label{eq:bc_va}
    A_v(j_f) = \left(\prod_{f}(2j_f + 1)^k\right) \tenjsymbol,
\end{equation}
where $\tenjsymbol$ denotes the Riemannian 10$j$ symbol and $k$ an integer which parametrises the choice of triangle weight in the measure\footnote{the SF formalism suggests using the measure that gives exactly a topologically invariant partition function before imposing the constraints reducing $BF$-theory to GR, which amount to choosing $k = 2$.}. The Riemannian 10$j$ symbol is a function of 10 spins obtained by evaluating the following closed SU(2)$\times$SU(2) spin-network on a flat connection \cite{Baez:2001et}
\begin{equation}
\begin{tikzpicture}[
    scale = 2.0,
    baseline = {(0,0)}, 
    rotate = 90, 
    every node/.style = {
        font = \footnotesize,
        inner sep = 1pt,
        minimum size = 5pt
    },
    rep/.style = {
        midway,
        fill = white,
        inner sep = 1pt
    }
]

  \foreach \i in {0, ..., 4}{
    \coordinate (v\i) at ({72*\i}:1);
    \node[draw, circle, minimum size = 4pt, fill] at (v\i) {};
  }
  
  \draw (v0) -- node[rep] {$j_{2}$} (v1);
  \draw (v1) -- node[rep] {$j_{3}$} (v2);
  \draw (v2) -- node[rep] {$j_{4}$} (v3);
  \draw (v3) -- node[rep] {$j_{5}$} (v4);
  \draw (v4) -- node[rep] {$j_{1}$} (v0);
  \draw (v0) -- node[rep] {$j_{7}$} (v2);
  \draw (v1) -- node[rep] {$j_{8}$} (v3);
  \draw (v2) -- node[rep] {$j_{9}$} (v4);
  \draw (v3) -- node[rep] {$j_{10}$} (v0);
  \draw (v4) -- node[rep] {$j_{6}$} (v1);
\end{tikzpicture}
\end{equation}
where the vertices are labelled by Barrett-Crane intertwiners. One can also define a modified 10$j$ symbol using the modified Barrett-Crane intertwiners as the weighted sum of SU(2) spin-networks \cite{Baez:2001et}. Since the value of a disjoint union of spin-networks is simply the product of their values, it follows that the modified 10$j$ symbol is \cite{Baez:2001et}
\begin{equation}
\label{eq:modified_10j_symbol}
\begin{tikzpicture}[
    scale = 1.8,
    baseline = {(0,0)},
    rotate = 90,
    every node/.style = {
        font = \footnotesize,
        inner sep = 1pt,
        minimum size = 5pt
    },
    rep/.style = {
        midway,
        fill = white,
        inner sep = 1pt
    }
]
  \foreach \i in {0, ..., 4}{
    \coordinate (v\i) at ({72*\i}:1);
    \draw[dotted, line width = 0.75pt] ({72*\i}:0.8) -- ({72*\i}:1.15);
    \node[draw, circle, minimum size = 4pt, fill] at (v\i) {};
  }
  \draw (v0) -- node[rep] {$j_{2}$} (v1);
  \draw (v1) -- node[rep] {$j_{3}$} (v2);
  \draw (v2) -- node[rep] {$j_{4}$} (v3);
  \draw (v3) -- node[rep] {$j_{5}$} (v4);
  \draw (v4) -- node[rep] {$j_{1}$} (v0);
  \draw (v0) -- node[rep] {$j_{7}$} (v2);
  \draw (v1) -- node[rep] {$j_{8}$} (v3);
  \draw (v2) -- node[rep] {$j_{9}$} (v4);
  \draw (v3) -- node[rep] {$j_{10}$} (v0);
  \draw (v4) -- node[rep] {$j_{6}$} (v1);
\end{tikzpicture}
\quad = \quad
\sum_{k_1,\ldots,k_5} \left( \prod_{i=1}^5 (2k_i + 1) \right)
\quad
\left(
\begin{tikzpicture}[
    scale = 1.8,
    baseline = {(0,0)},
    every node/.style = {
        font = \scriptsize
    },
    lab/.style = {
        midway,
        fill = white,
        inner sep = 1pt
    },
    dot/.style = {
        draw,
        circle,
        fill,
        minimum size = 4pt,
        inner sep = 0pt
    }
]
  \foreach \i [evaluate=\i as \ang using 36*\i] in {0,...,9}{
    \coordinate (v\i) at (\ang:1);
    \node[dot] at (v\i) {};
  }
  \draw (v0) -- node[lab] {\(k_{1}\)} (v1);
  \draw (v1) -- node[lab] {\(j_{1}\)} (v2);
  \draw (v2) -- node[lab] {\(k_{2}\)} (v3);
  \draw (v3) -- node[lab] {\(j_{2}\)} (v4);
  \draw (v4) -- node[lab] {\(k_{3}\)} (v5);
  \draw (v5) -- node[lab] {\(j_{3}\)} (v6);
  \draw (v6) -- node[lab] {\(k_{4}\)} (v7);
  \draw (v7) -- node[lab] {\(j_{4}\)} (v8);
  \draw (v8) -- node[lab] {\(k_{5}\)} (v9);
  \draw (v9) -- node[lab] {\(j_{5}\)} (v0);
  \draw (v1) -- node[lab] {\(j_{6}\)} (v4);
  \draw (v2) -- node[lab] {\(j_{10}\)} (v9);
  \draw (v3) -- node[lab] {\(j_{7}\)} (v6);
  \draw (v5) -- node[lab] {\(j_{8}\)} (v8);
  \draw (v7) -- node[lab] {\(j_{9}\)} (v0);
\end{tikzpicture}
\right)^2
\end{equation}
Given the choice of splitting shown above, the modified 10$j$ symbol can be shown to be always non-negative and is related to the 10$j$ symbol by \cite{Baez:2001et}
\begin{equation}
\label{eq:10j_to_modified_10j}
\begin{tikzpicture}[
    scale = 2.0,
    baseline = {(0,0)}, 
    rotate = 90, 
    every node/.style = {
        font = \footnotesize,
        inner sep = 1pt,
        minimum size = 5pt
    },
    rep/.style = {
        midway,
        fill = white,
        inner sep = 1pt
    }
]

  \foreach \i in {0, ..., 4}{
    \coordinate (v\i) at ({72*\i}:1);
    \node[draw, circle, minimum size = 4pt, fill] at (v\i) {};
  }
  
  \draw (v0) -- node[rep] {$j_{2}$} (v1);
  \draw (v1) -- node[rep] {$j_{3}$} (v2);
  \draw (v2) -- node[rep] {$j_{4}$} (v3);
  \draw (v3) -- node[rep] {$j_{5}$} (v4);
  \draw (v4) -- node[rep] {$j_{1}$} (v0);
  \draw (v0) -- node[rep] {$j_{7}$} (v2);
  \draw (v1) -- node[rep] {$j_{8}$} (v3);
  \draw (v2) -- node[rep] {$j_{9}$} (v4);
  \draw (v3) -- node[rep] {$j_{10}$} (v0);
  \draw (v4) -- node[rep] {$j_{6}$} (v1);
\end{tikzpicture}
\quad = \quad
(-1)^{2(j_1 + \cdots + j_{10})}
\quad
\begin{tikzpicture}[
    scale = 2.0,
    baseline = {(0,0)},
    rotate = 90,
    every node/.style = {
        font = \footnotesize,
        inner sep = 1pt,
        minimum size = 5pt
    },
    rep/.style = {
        midway,
        fill = white,
        inner sep = 1pt
    }
]
  \foreach \i in {0, ..., 4}{
    \coordinate (v\i) at ({72*\i}:1);
    \draw[dotted, line width = 0.75pt] ({72*\i}:0.8) -- ({72*\i}:1.15);
    \node[draw, circle, minimum size = 4pt, fill] at (v\i) {};
  }
  \draw (v0) -- node[rep] {$j_{2}$} (v1);
  \draw (v1) -- node[rep] {$j_{3}$} (v2);
  \draw (v2) -- node[rep] {$j_{4}$} (v3);
  \draw (v3) -- node[rep] {$j_{5}$} (v4);
  \draw (v4) -- node[rep] {$j_{1}$} (v0);
  \draw (v0) -- node[rep] {$j_{7}$} (v2);
  \draw (v1) -- node[rep] {$j_{8}$} (v3);
  \draw (v2) -- node[rep] {$j_{9}$} (v4);
  \draw (v3) -- node[rep] {$j_{10}$} (v0);
  \draw (v4) -- node[rep] {$j_{6}$} (v1);
\end{tikzpicture}
\end{equation}
Given that $\tenjsymbol$ is the vertex amplitude in this model, several algebraic and numerical algorithms \cite{Christensen:2001eu,Baez:2002rx,Baez:2002aw} for evaluating $\tenjsymbol$, as well as its large spin asymptotics, have been extensively developed and studied, yet the computational cost still scales unfavourably with the magnitudes of the spins.
\newparagraph
In this work, we will explore a data-driven alternative: casting the evaluation of the 10$j$ symbol, and thus the vertex amplitude of the Barrett-Crane model, as a supervised learning problem for deep neural networks. Concretely, we (i) generate a comprehensive training set of exact 10$j$ symbol values using an optimised algorithm, (ii) break down the learning task into classification and regression tasks and (iii) quantify the fidelity and generalisation of the learned amplitude on spins that lie outside the training domain. The goals of this work are twofold. First, we demonstrate a \emph{proof-of-principle}: (high dimensional) vertex amplitudes of spinfoam models are amenable to modern deep learning techniques. Second, we lay the groundwork for accelerating numerical investigations of more realistic models. Because the EPRL/FK vertex reduces to its BC counterpart in specific limits, the aim is for the models and tools used and developed in this work to be ported, with appropriate modifications capturing the $\gamma$-dependence, to the state-of-the-art SF models. 
\newparagraph
The presentation of the work is as follows:
\begin{itemize}
    \item[(i)] In Section \ref{sec:methodology}, we start by clearly stating the goals of the current work as well as presenting the methodology to be used.
    \item[(ii)] In Section \ref{sec:data_gen_and_process}, we discuss the generation of training data and the pre-processing of the produced datasets prior to the training process.
    \item[(iii)] Section \ref{sec:net_arch_and_enc} discusses the specific architecture of the networks used in this work and encoding schemes used to facilitate the learning process.
    \item[(iv)] Section \ref{sec:protocols} concerns the training protocols for both the classification and regression tasks conducted in this work as well as the evaluation metrics to which we evaluate the networks post-training.
    \item[(v)] In Section \ref{sec:results}, we present the results of the training for both, classification and regression tasks, and
    \item[(vi)] lastly, in Section \ref{sec:discussion} we go over the issues encountered in this work, the limitations of the current work and how it relates to future models to be considered and present some avenues for exploration to said problems.
\end{itemize}

\section{Methodology}
\label{sec:methodology}
As mentioned, this study serves as a proof-of-principle to address the question of whether or not a neural network be utilised to learn the vertex amplitude of a given spinfoam model, in this case the Euclidean Barrett-Crane model. The approach we use is one of supervised learning  (SL). We approach the problem in terms of two tasks, which can be roughly outlined as follows. First, a classification task where, given a data set $\dataset{S}{A(\mathbf{S})}$ of spin configurations $\mathbf{S}$ and corresponding amplitudes $A(\mathbf{S})$, we train a classifer $\classifier(\mathbf{S})$ to determine whether the corresponding amplitude $A(\mathbf{S})$ is zero or not. Next, a regression task where once again, given the dataset described above, we train a regressor $\regressor(\mathbf{S})$ to predict the correct amplitude $A(\mathbf{S})$ for the given spin configuration $\mathbf{S}$. We then construct a meta ``Expert" network $\meta(\mathbf{S})$ which combines both $\classifier(\mathbf{S})$ and $\regressor(\mathbf{S})$ to provide the correct predicted amplitude for the given configuration $\mathbf{S}$.
\newparagraph
We will proceed by detailing the components of the implementation in natural order. That is, we first briefly describe data acquisition and pre-processing followed by the network architectures and encoding schemes. Next, we outline the training protocol, hyper-parameter choices and monitored metrics. Following that, for each task we present the results for the mentioned evaluation metrics. A discussion regarding limitations, technical details and ablation studies are conducted in the Discussion (see Section \ref{sec:discussion}).

\subsection{Data generation and processing}
\label{sec:data_gen_and_process}
The core component of the vertex amplitude in the Euclidean Barrett-Crane model is the $\tenjsymbol$, which can either be negative, positive or zero. To simplify our task, we will focus on learning the square of the $\tenjsymbol$, for both classification and regression. For classification, the appropriate sign factor can be easily reconstructed from the given spin configuration. This effectively allows us to reduce the complexity of the classification task while remain able to reconstruct the correct sign of the learned $\tenjsymbol$. For regression, the square root of the prediction can be taken at inference. The tools and software used in this work are all Python based. As such, to facilitate the training process, the algorithm presented in \cite{Christensen:2001eu} and the corresponding implementation in C provided in therein\footnote{\texttt{http://jdc.math.uwo.ca/spin-foams/10j-code/}, (accessed April 2025)} has been rewritten in Python and accelerated by utilising just-in-time compilation using \code{numba} \cite{LLVM:CGO04} to obtain a comparatively fast compile time compared to the C implementation. 
\newparagraph
For both tasks, we respectively train the networks within a specified spin cutoff. That is, for a given cutoff spin $j_{max}$, then the spin configurations to be considered are ones such that $(j_1, j_2, \cdots, j_{10}) =: \mathbf{S^{j_{max}}} \in (\mathcal{S}^{(j_{max})})^{10}$ where $\mathcal{S}^{(j_{max})} = \left\lbrace j \in \frac{1}{2}\mathbb{Z} \vert j \leq j_{max} \right\rbrace$, $|\mathcal{S}^{(j_{max})}| = K := 2j_{max} + 1$ and the superscript is explicitly stated in $\mathbf{S^{(j_{max})}}$ to indicate the cutoff to which the configuration belongs to. The following datasets are created:
\begin{itemize}
    \item[(i)] For classification: a dataset $\mathcal{D}_\classifier^{(j_{max})} := \dataset{S^{(j_{max})}}{\sigma(\mathbf{S^{(j_{max})}})}$ containing pairs of spin configurations $\mathbf{S^{(j_{max})}}$ and their corresponding $\tenjsymbol^2$ signs $\sigma (\mathbf{S^{(j_{max})}}) := \sign (\tenjsymbol(\mathbf{S^{(j_{max})}}))^2$. Since $(\tenjsymbol (\mathbf{S^{(j_{max})}}))^2 \in \mathbb{R}_{\geq 0}$, then for any $\mathbf{S^{(j_{max})}}$ it follows that $\sigma(\mathbf{S^{(j_{max})}}) \in [0, 1]$. This dataset contains data points for \emph{all} possible configurations $\mathbf{S^{(j_{max})}}$ in a given cutoff.

    \item[(ii)] For regression: another dataset $\mathcal{D}_\regressor^{(j_{max})} := \dataset{S^{(j_{max})}}{\logtenj{\mathbf{S^{(j_{max})}}}}$ which then contains pairs of spin configurations $\mathbf{S^{(j_{max})}}$ and their corresponding $\logtenj{\mathbf{S^{(j_{max})}}} := \log\left((\tenjsymbol(\mathbf{S^{(j_{max})}}))^2 + \epsilon\right)$ where $\epsilon = 1\times 10^{-26}$ is a small correction factor. Unlike $\mathcal{D}_\classifier^{(j_{max})}$, this dataset contains data points for all configurations $\mathbf{S^{(j_{max})}}$ in a given cutoff \emph{which have a non-zero} $\logtenj{\mathbf{S^{(j_{max})}}}$ \emph{value}. Thus, $|\mathcal{D}_\regressor^{(j_{max})}| < |\mathcal{D}_\classifier^{(j_{max})}|$ always.
\end{itemize}
For small enough cutoffs, such datasets can be obtained simply by full enumeration of the space $\mathcal{S}^{(j_{max})}$. To understand the distribution of the potential training data, one can compare the number of non-zero $\logtenj{\mathbf{S^{(j_{max})}}}$ for all $\mathbf{S^{(j_{max})}}$ in different cutoffs.
\begin{figure}[h]
    \centering
    \includegraphics[width=0.6\linewidth]{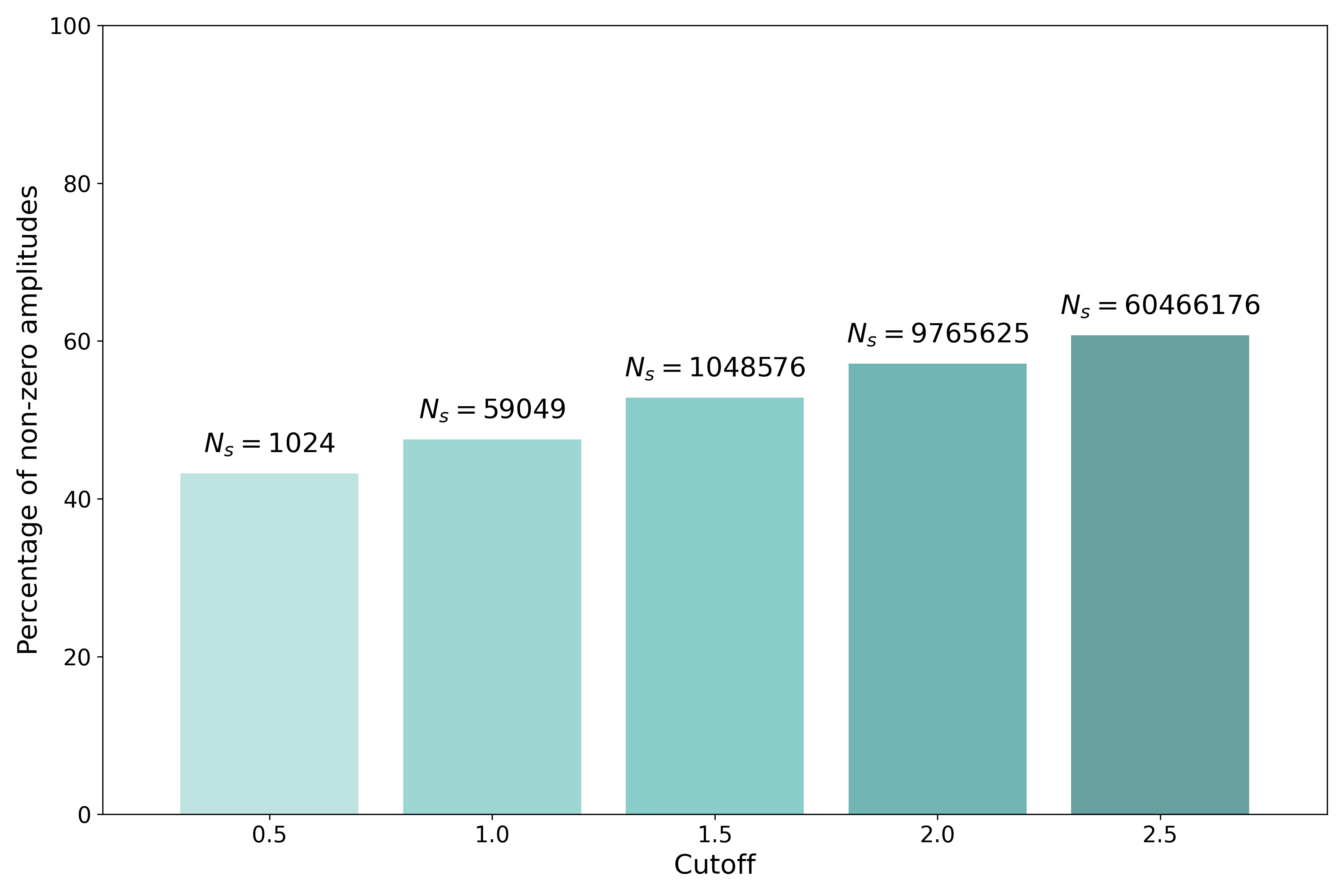}
    \caption{Percentage of non-zero to zero vertex amplitudes at different cutoffs with the total number of states $N_s$ shown for each cutoff.}
    \label{fig:percentage_nnz_amps}
\end{figure}
\\
It is immediately evident then, as shown in Figure \ref{fig:percentage_nnz_amps}, that there is an imbalance in the number of configurations which yield a zero amplitude as compared to non-zero amplitude. In fact, as the cutoff increases, the percentage of non-zero amplitude configurations increases drastically, reaching over 60\% already at $j_{max} = 2.5$. Further, the value of $(\tenjsymbol(\mathbf{S^{(j_{max})}}))^2$ can vary by more than 22 orders of magnitude in even small cutoffs as $j_{max} = 2.0$. This therefore poses the following hurdles: (i) a class imbalance for classification and (ii) very large range in the values of $\logtenj{\mathbf{S^{(j_{max})}}}$ for the regression.

\subsection{Network architecture and encoding schemes}
\label{sec:net_arch_and_enc}
As this study is a proof-of-principle, little emphasis was put on constructing a sophisticated network architecture for either task. Remarkably, as will be shown, simple architectures suffice for both tasks. A multi-layer perceptron (MLP) \cite{Rumelhart:1986gxv,Hornik:1989yye} was used for both the classification and regression tasks. For classification, the classifier $\classifier(\mathbf{S})$ consisted of a MLP with one hidden layer (depth 1) and 128 hidden nodes (width 128). A rectified linear unit (ReLU) activation function 
\begin{equation}
    \mathrm{ReLU}(x) = \max(0, x)
\end{equation}
was used to introduce the non-linearity. The input for the classifier $\classifier(\mathbf{S})$ was simply the spin configuration $\mathbf{S^{(j_{max})}}$. The architecture for the classifier remained constant irrespective of the training cutoff. The classifier maintained a number of trainable parameters $\texttt{Params}(\classifier)$ of only 1537 parameters.
\newparagraph
For the regressor $\regressor(\mathbf{S})$, a MLP was used as well. Unlike $\classifier(\mathbf{S})$, the depth and width of $\regressor(\mathbf{S})$ varied depending on the $j_{max}$ chosen during training. For example, at $j_{max} = 1.0$, $\regressor(\mathbf{S})$ had a depth of 6 and a width of 256 while for $j_{max} = 1.5$, it had a depth of 5 and a width of 512. Different activation functions were considered and tested. Ultimately, a Gaussian error linear units activation function
\begin{equation}
    \mathrm{GELU}(x) = x\Phi(x),
\end{equation}
where $\Phi(x)$ is the cumulative distribution function for Gaussian distribution, was used. Unlike ReLU activation, the derivative of the GELU function is continuous at the origin. A smooth activation allows the optimiser to track higher-order curvature and avoids the kinks that piece‑wise linear units, such as ReLU, introduce. Further, while ReLU either passes the signal unchanged ($x$) or blocks it (0), the GELU multiplies the input $x$ by the probability that a unit drawn from a standard Gaussian is below $x$. Small inputs are only tempered, not annihilated, which preserves information carried by low‑magnitude features (inputs).
\newparagraph
While the classifier $\classifier(\mathbf{S})$ takes as an input the configuration $\mathbf{S}$, the regressor $\regressor(\mathbf{S})$ does not. The input for $\regressor(\mathbf{S})$ is encoded using an indexing function $\Omega$. For a single $j \in \mathcal{S}^{(j_{max})}$, let $\mathbf{i}: \mathcal{S}^{(j_{max})} \rightarrow \left\lbrace 0, 1, \cdots, K - 1\right\rbrace$ such that $\mathbf{i}(j) = 2j$. Let $\Vec{e}_r \in \mathbb{R}^K$ be the standard $r$-th basis vector such that $(\Vec{e}_r)_q = \delta_{q, r}$. We define the one-hot encoding of a single spin to be the map $\omega: \mathcal{S}^{(j_{max})} \rightarrow \left\lbrace 0, 1 \right\rbrace^K \subset \mathbb{R}^K$ such that $\omega(j) = \Vec{e}_{\mathbf{i}(j)}$ which places a 1 exactly at the coordinate that corresponds to the value of $j$. The one-hot encoding for a configuration $\mathbf{S^{j_{max}}}$ is then $\Omega : (\mathcal{S}^{(j_{max})})^{10} \rightarrow \left\lbrace 0, 1 \right\rbrace^{10K} \subset \mathbb{R}^{10K}$. Simply put, while the classifier takes as an input a configuration of 10 values in a given $j_{max}$, the regressor takes as an input $10K$ values at the same $j_{max}$. The reason for this is that spin values merely are categorical symbols, not points on a physically meaningful real line (e.g. $j = 1$ is not ``twice as large" as $j = 1/2$ in the sense that Euclidean distance would suggest). Consequently, we convert every spin to a pure indicator vector rather than feed its numeric value directly. This removes artificial ordinality and provides compatibility with transfer learning across cutoffs. Note that one can in principle choose any faithful encoding map, not necessarily restricted to the one mentioned above.

\subsection{Training protocol and hyperparameter choices}
\label{sec:protocols}
This work utilises PyTorch \cite{Paszke:2019xhz} for all neural network related aspects. A non-exhaustive preliminary assessment conducted via automatic hyperparameter tuning using \texttt{optuna} \cite{Akiba:2019lwq} was carried out to determine an estimate for the best hyperparameters in this work. The following parameters, shared for both tasks, were thus obtained. The networks in both tasks were trained with a mini-batch AdamW optimiser \cite{Loshchilov:2017bsp} with a weight decay of $1\times 10^{-6}$ and a 1-cycle learning rate schedule with a peak step size of $1 \times 10^{-3}$ for the case of the regressor and $1 \times 10^{-4}$ for the case of the classifier. The batch size for the classification task was 8 while a batch size of 256 was chosen for the regression task. For both tasks, training takes place in the low spin regime, as the behvaiour of the network evaluations can be extrapolated to further training on higher spins. In what follows, we outline task specific evaluation metrics and further training protocols.

\subsubsection{Classification metrics and protocol}
\label{sec:class_metrics_prot}

In what follows, we often refer to the $\sigma(\mathbf{S^{(j_{max})}})$ for a given $\mathbf{S^{(j_{max})}}$ as a \emph{label}. In this binary classification task, that label can either be 0 or 1. The binary classification task was optimised with respect to a weighted binary cross-entropy loss function 
\begin{equation}
    \mathrm{BCELoss} = -\frac{1}{N}\sum_{i = 1}^{N} \left[ \omega_1 y_i \log p_i + \omega_0 (1 - y_i)\log (1 - p_i) \right],
\end{equation}
as standard for such classification tasks, where $(\omega_0, \omega_1)$ are class weights used to neutralise the imbalance between zero and non-zero amplitudes, $y_i$ denotes the actual binary label (0 or 1) of the $i$-th observation, $p_i$ denotes the probability of the $i$-th observation to be in the class 1 and $N$ is the total number of observations made. 
\newparagraph
The training protocol is done cutoff-wise. First, a network is trained and evaluated at a cutoff of 0.5. Since the network architecture is static between cutoffs, transfer learning was utilised and the network was then retrained, starting with previously optimised parameters from the preceding cutoff, on a cutoff of 1.0 and then once again evaluated. This protocol continued until a cutoff of 2.0. The training dataset size starts at 75\% of all available configurations at a cutoff of 0.5. As transfer learning was conducted successively for higher cutoffs, the training dataset size was decreased incrementally to establish a minimal dataset size: the smallest dataset size required to achieve the highest evaluation metrics values. Note that the datasets were collected blindly, no active binning in terms of cutoffs or have been employed. While this is possible, it has not been done intentionally to test the model's capacity to train from blind data.
\newparagraph
Several metrics were evaluated after each cutoff training cycle. Those are
\begin{itemize}
    \item[(i)] Hard accuracy: this is defined as
    \begin{equation}
        \mathrm{AccHard} = \frac{1}{N}\sum_{i = 1}^N 1 \cdot \left\lbrace \texttt{f}(\hat{y}_i) = y_i\right\rbrace,
    \end{equation}
    which essentially counts, on the test sample of size $N$, how many network predicted labels $\hat{y}_i$ exactly match the true labels $y_i$. Here, $\texttt{f}(\hat{y}_i)$ is a decision threshold function which returns 1 if $\hat{y}_i \geq 0.5$ and 0 otherwise.

    \item[(ii)] Soft accuracy: this is defined as
    \begin{equation}
        \mathrm{AccSoft} = \frac{1}{N} \sum_{i = 1}^N (1 - |\hat{y}_i - y_i|)
    \end{equation}
    which essentially computes how ``far off" was the network prediction from the true labels for a given test batch of size $N$.

    \item[(iii)] Precision \cite{Powers:2020}: which is defined as
    \begin{equation}
        P = \frac{TP}{TP + FP},
    \end{equation}
    where $TP$ denotes True Positives (data points with labels 1 which have been correctly predicted by the network) and $FP$ denotes False Positives (data points with labels 0 which have been incorrectly predicted by the network to have label 1), conveys the fraction of all predicted non-zero states that are correct. Essentially, this metric focuses on the quality of positive predictions, giving a measure of how trustworthy is the network's prediction when predicting a non-zero labels.

    \item[(iv)] Recall \cite{Powers:2020}: the ability to correctly predict all non-zero configurations which is defined as
    \begin{equation}
        R = \frac{TP}{TP + FN},
    \end{equation}
    where $FN$ denotes False Negatives (data points with labels 1 which have been incorrectly predicted by the network to have a label 0). Having a high recall value indicates that the network rarely fails to correctly label a non-zero configuration.

    \item[(v)] F-1 score \cite{Powers:2020}: defined as
    \begin{equation}
        \mathrm{F1} = 2 \frac{P \cdot R}{P + R},
    \end{equation}
    is the harmonic mean of precision and recall, symmetrically representing both precision and recall in one metric. The highest obtainable F-1 score of 1.0 indicates perfect precision and recall.
\end{itemize}
Additionally, for every cutoff training round, a confusion matrix, evaluated on the entire configuration space, is computed. This gives a clear picture of the number of FN and FP values for the network.

\subsubsection{Regression metrics and protocol}
\label{sec:reg_metrics_prot}

Unlike the classification task. The regressor was trained on a single chosen cutoff. No transfer learning was utilised. Therefore, a regressor $\regressor$ trained at a cutoff of 0.5 would only be evaluated on states $\mathbf{S^{(0.5)}}$. This is due to the dynamic nature of network's architecture used in this task which makes transfer learning, although not impossible, much harder. For all cutoffs, the training data consisted of 85\% of all available non-zero configurations $\mathbf{S^{(j_{max})}}$ at the current cutoff. Since the labels in this task can have a rather large range, the training dataset was not blindly collected. Rather, the entire space was first enumerated after which the amplitudes were binned according to their magnitude. The produced dataset of size 85\% of the all non-zero contributing data points was stratified according to those bins, ensuring sufficient representation across all magnitudes available.
\newparagraph
The loss function for this task was chosen to be the Huber loss \cite{Huber:1964}
\begin{equation}
    \mathrm{HuberLoss} (e) = \begin{cases}
        \frac{1}{2}e^2 &\quad,\qquad |e| < \delta \\
        \delta(|e| - \frac{1}{2}\delta) & \quad,\qquad |e| \geq \delta
    \end{cases}
\end{equation}
where $e := \widehat{\logtenj{\mathbf{S^{(j_{max})}}}} - \logtenj{\mathbf{S^{(j_{max})}}}$. For the case of $\delta = 1$, this is equivalent to the smooth L1 loss function. Here, $\widehat{\logtenj{\mathbf{S^{(j_{max})}}}}$ denotes the network predicted log value of the square of the $\tenjsymbol$ symbol of the given configuration. For brevity, we will denote that with $\hat{y}_{\log}$ and denote the true value with $y_{\log}$. For any training cutoff, the following metrics were observed after training:
\begin{itemize}
    \item[(i)] Root mean squared error (RMSE) in log space: this is simply defined as $\mathrm{RMSE}_{\log} = \sqrt{\mathrm{MSE}_{\log}}$ where
    \begin{equation}
        \mathrm{MSE}_{\log} = \frac{1}{N} \sum_{i = 1}^N (\hat{y}_{\log} - y_{\log})^2.
    \end{equation}
    Note that this metric may be sensitive to errors in large valued labels. Applying it in log space is due to the large possible label magnitudes which span several orders. This then ensures that large values do not disproportionately dominate the error.

    \item[(ii)] Median absolute deviation (MAD) in log space: defined as
    \begin{equation}
        \mathrm{MAD}_{\log} = \mathrm{Median}\left(\left\lbrace |\hat{y}_{\log}^{(i)} - y_{\log}^{(i)}| \right\rbrace_{i = 1}^{N} \right)
    \end{equation}
    is a measure of error also in the log space. Unlike the $\mathrm{RMSE}_{\log}$, this metric is resilient to outliers and better reflects the typical prediction deviation.

    \item[(iii)] Mean absolute percentage error (MAPE): which is given as
    \begin{equation}
        \mathrm{MAPE} = \frac{100}{N}\sum_{i = 1}^N \left\lvert \frac{e^{\hat{y}_{\log}} - e^{y_{\log}}}{e^{y_{\log}}} \right\rvert,
    \end{equation}
    which can be interpreted in terms of relative error of predictions over a test set of size $N$. Note that this metric is not computed in log space.

    \item[(iv)] Threshold accuracy: lastly, we measure the threshold accuracy which is
    \begin{equation}
        \mathrm{Acc}_{\leq\epsilon} = \frac{1}{N} \sum_{i = 1}^N 1\cdot \left\lbrace |e^{\hat{y}_{\log}} - e^{y_{\log}}| \leq \epsilon |e^{y_{\log}}|\right\rbrace
    \end{equation}
    which is an indicator of how many predictions in a test set of size $N$ have a relative error lower than a specified $\epsilon$ threshold. In this work, we take $\epsilon = 0.1$ and thus the $\mathrm{Acc}_\epsilon$ will measure how many predictions fall within a 10\% relative error to the true value. Once again, this metric is not computed in log space.
\end{itemize}
Additionally, we also compute the $R^2$ value for as well as a true vs. prediction plot after every cutoff training cycle. 

\section{Training results}
\label{sec:results}
\begin{figure}[h]
    \centering
    \includegraphics[width=0.9\linewidth]{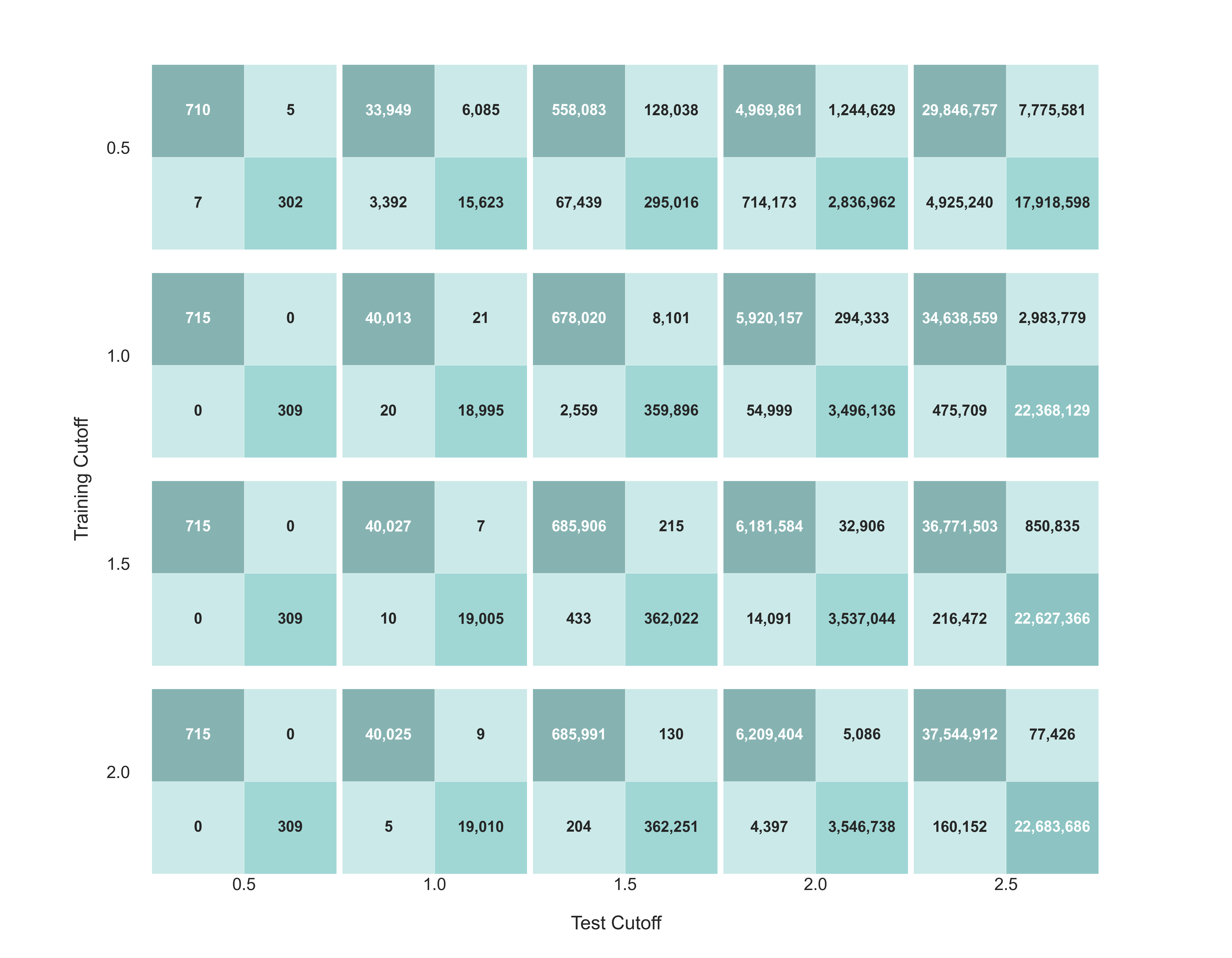}
    \caption{The confusion matrices for the classifier at different stages of training tested on different cutoffs each time after each training round.}
    \label{fig:conf_mat}
\end{figure}

\begin{figure}[h]
    \centering
    \begin{subfigure}[t]{0.495\textwidth}
        \centering
        \includegraphics[width=\textwidth]{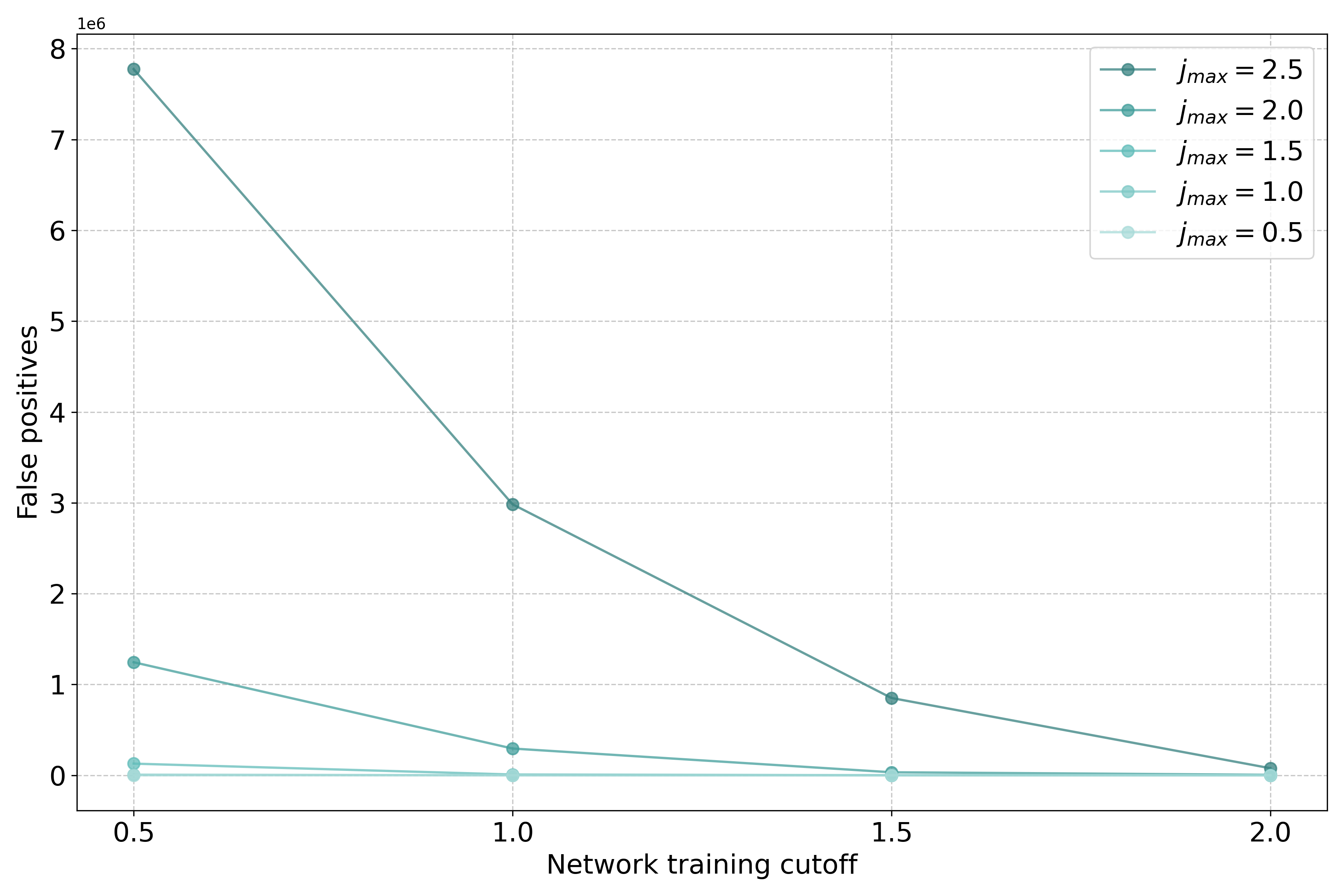}
        \caption{}
        \label{fig:subfig_FP}
    \end{subfigure}
    \hfill
    \begin{subfigure}[t]{0.495\textwidth}
        \centering
        \includegraphics[width=\textwidth]{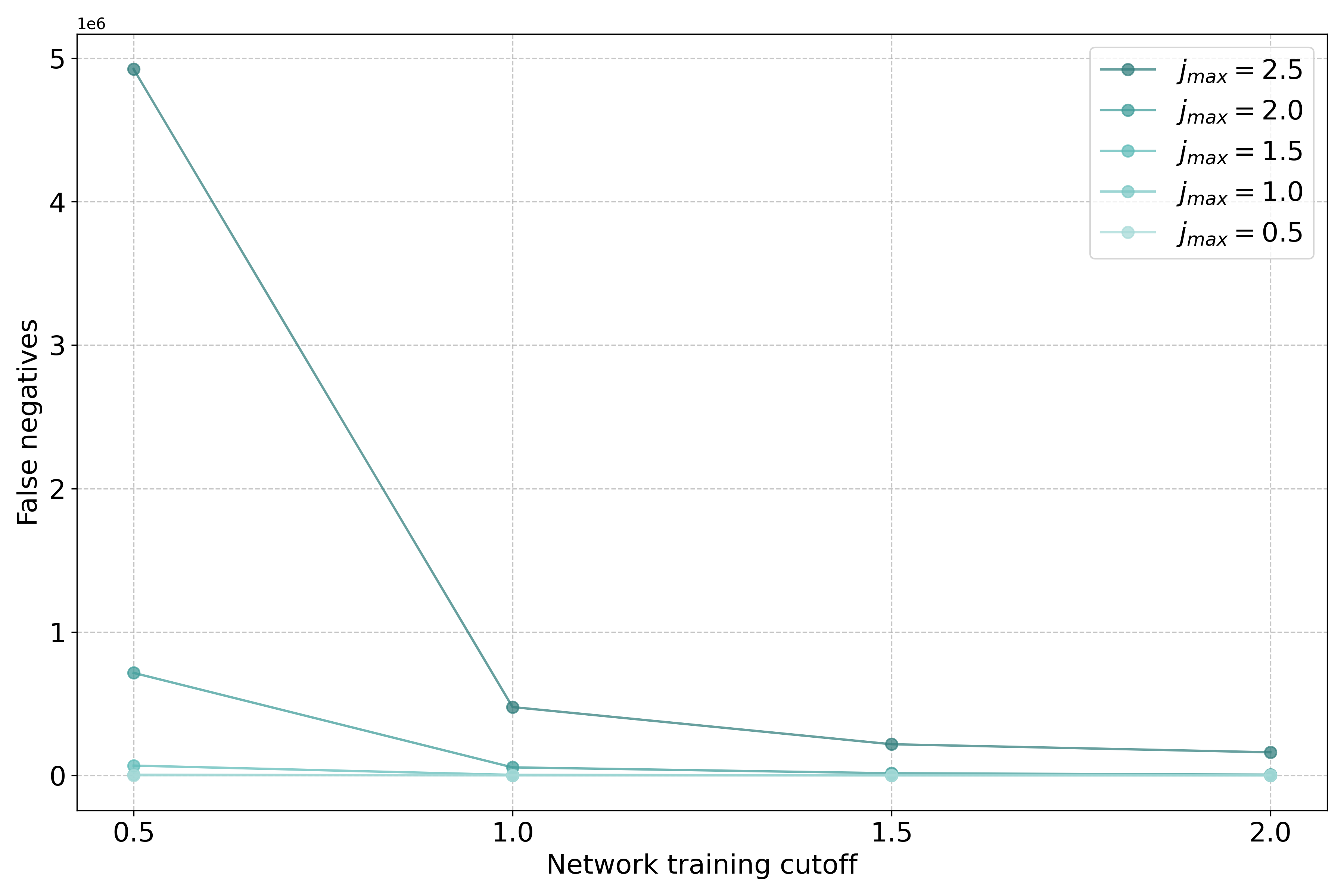}
        \caption{}
        \label{fig:subfig_FN}
    \end{subfigure}
    \caption{The number of false positives and false negatives for the classifier trained at different cutoffs and tested on cutoffs from 0.5 to 2.5 is shown on the left and right respectively.}
    \label{fig:FP_FN}
\end{figure}

The two criteria being sought in this work for both tasks are (i) if the networks can perform well at the cutoff they are trained in and (ii) if the networks can predict on test samples from a cutoff they have not been trained on. All computations were carried out on an Intel Xeon E3-1240 v5 with 4 cores at 3.5GHz and no distributed, parallelised or GPU computations were utilised. We begin with the classification task. Carrying out the protocol described in Section \ref{sec:class_metrics_prot}, the results shown in Table \ref{tab:class_loss_results} were observed.
\begin{table}[h]
    \centering
    \begin{tabular}{ccccccc}
        \rowcolor{palegreen!50}
        $j_{max}$ & Training Loss & Test Loss & Training Time (sec) & $N_s$ & $N_{\mathrm{train}}$ & $N_{\mathrm{train}} / N_s$ \\
        \hline
        0.5 & 0.01772 & 0.15405 & 59.7 & 1024 & 768 & 0.75 \\
        1.0 & $8.5601\times10^{-7}$ & 0.0362 & 2807.28 & 59049 & 36126 & 0.611 \\
        1.5 & $1.49516\times10^{-7}$ & 0.01092 & 13386.76 & 1048576 & 103618 & 0.098 \\
        2.0 & 0.00236 & 0.02397 & 24134.8 & 9765625 & 170356 & 0.017 \\
    \end{tabular}
    \caption{The training and test loss for the classification task on different cutoffs is shown. Here, $N_s$ denotes the total number of possible configurations at the cutoff and $N_{\mathrm{train}}$ denotes the number of samples used in the training process.}
    \label{tab:class_loss_results}
\end{table}
\\
Table \ref{tab:class_loss_results} shows the training and test loss for the classification task on different cutoffs. As shown, the training dataset size starts at 75\% of the total number of available configurations at the cutoff $N_s$ for $j_{max} = 0.5$. As transfer learning is applied and training proceeded to higher cutoffs, the training dataset size decreases until 1.17\% for $j_{max} = 2.0$. Despite that, both training and test loss decrease steadily as the cutoff increases, indicating that the learning process is carried out successfully. The increase in the test and training loss for $j_{max} = 2$ is attributed to the relatively small training dataset size. 
\newparagraph
After each training cycle, the trained network was tested on all configurations for cutoffs $j_{max} \in [0.5, 1.0, 1.5, 2.0, 2.5]$. The following metrics in Table \ref{tab:class_cutoff_metrics} were observed. 
\begin{table}[h]
    \centering
    \begin{tabular}{cccccccc}
        \rowcolor{palegreen!50}
        Training $j_{max}$ & Test $j_{max}$ & AccSoft (\%) & AccHard (\%) & $P$ & $R$ & F1 \\
        \hline
        \multirow{5}{*}{0.5} & 0.5 & 96.7432 & 98.8281 & 0.9837 & 0.9773 & 0.9804 \\
                            & 1.0 & 83.3652 & 83.9506 & 0.7197 & 0.8216 & 0.7672 \\
                            & 1.5 & 81.0237 & 81.3578 & 0.6973 & 0.8139 & 0.7511 \\
                            & 2.0 & 79.7331 & 79.9418 & 0.6951 & 0.7989 & 0.7433 \\
                            & 2.5 & 78.8568 & 78.9951 & 0.6974 & 0.7844 & 0.7383 \\
        \hline
        \multirow{5}{*}{1.0} & 0.5 & 99.9996 & 100 & 1.0 & 1.0 & 1.0 \\
                            & 1.0 & 99.9392 & 99.9395 & 0.9989 & 0.9989 & 0.9989 \\
                            & 1.5 & 98.9591 & 98.9833 & 0.9780 & 0.9929 & 0.9853 \\
                            & 2.0 & 96.3990 & 96.4228 & 0.9223 & 0.9845 & 0.9523 \\
                            & 2.5 & 94.2651 & 94.2786 & 0.8823 & 0.8792 & 0.8792 \\
        \hline
        \multirow{5}{*}{1.5} & 0.5 & 99.9999 & 100 & 1.0 & 1.0 & 1.0 \\
                            & 1.0 & 99.9719 & 99.9712 & 0.9996 & 0.9995 & 0.9995 \\
                            & 1.5 & 99.9369 & 99.9382 & 0.9994 & 0.9988 & 0.9990 \\
                            & 2.0 & 99.5092 & 99.5187 & 0.9908 & 0.9960 & 0.9933 \\
                            & 2.5 & 98.2248 & 98.2348 & 0.9638 & 0.9905 & 0.9769 \\
        \hline
        \multirow{5}{*}{2.0} & 0.5 & 99.9994 & 100 & 1.0 & 1.0 & 1.0 \\
                            & 1.0 & 99.9773 & 99.9762 & 0.9995 & 0.9997 & 0.9995 \\
                            & 1.5 & 99.9666 & 99.9681 & 0.9996 & 0.9994 & 0.9994 \\
                            & 2.0 & 99.8993 & 99.9028 & 0.9986 & 0.9988 & 0.9986 \\
                            & 2.5 & 99.5982 & 99.6070 & 0.9966 & 0.9930 & 0.9947 \\
        \hline
    \end{tabular}
    \caption{Classification metrics for the classifier trained with transfer learning at different cutoffs.}
    \label{tab:class_cutoff_metrics}
\end{table}
\newparagraph
As shown in Table \ref{tab:class_cutoff_metrics}, all classification metrics are within excellent range for the cutoff that the network has been trained on. However, one striking aspect is that the simple classifier seems to be able to perform reasonably well when tested on higher cutoffs which it has not been trained on. Additionally, no catastrophic forgetting\footnote{the classifier forgetting about previously learned cutoffs after being trained on a higher cutoff} was observed. Further, as the classifier is trained on higher cutoff, using transfer learning, the metrics for test cutoffs which fall above the training cutoff also improve drastically. For example, a classifier trained at $j_{max} = 0.5$ and shows an F-1 score of 0.7383 when tested on $j_{max} = 2.5$ while the same classifier retrained using transfer learning up to $j_{max} = 2.0$ has an F-1 score of 0.9947 for a test cutoff of $j_{max} = 2.5$. This can be further elucidated by looking at the confusion matrices for the classifier at different training stages as shown in Figure \ref{fig:conf_mat}.
\newparagraph
Figure \ref{fig:conf_mat} shows the confusion matrices for the classifier at different stages of training tested on different cutoffs at each stage. In a given confusion matrix, the top left quadrant denotes the the true negatives, the top right quadrant denotes the false negatives, the bottom left quadrant denotes the false positives and the bottom right quadrant denotes the true positives. It is evident that as the classifier progresses in the training stages, the number of false predictions, in either class, become lower. This can be easier shown by looking at only the false negatives and false positives as shown in Figure \ref{fig:FP_FN}.
\newparagraph
As can be seen in Figure \ref{fig:FP_FN}, both the false positives and false negatives fall drastically. Perhaps most interestingly, the number of false negatives and false positives is seen to decrease even on cutoffs on which the classifier has not been trained on (see the trend-line of $j_{max} = 2.5$ in both figures), as the classifier training progresses to higher cutoffs. The takeaway here is that despite the simple architecture of the classifier and the extremely small size of training datasets for higher cutoffs, the classifier is able to generalise well to cutoffs beyond its training.
\newparagraph
As discussed in the protocol in Section \ref{sec:reg_metrics_prot}, we approach the regression task differently from the classification. In what follows, we mainly focus on a regressor trained at cutoffs of 1.0 and 1.5. In the discussion section (Section \ref{sec:discussion}), we elaborate on higher cutoffs. 
\newparagraph
A regressor with a depth of 6 and width of 256 with a GELU activation was used for training at $j_{max} = 1.0$ while for $j_{max} = 1.5$, the regressor had a depth and width of 5 and 512 respectively. In both cases, training was conducted over 200 epochs and the training dataset consisted of 85\% of all configurations which yield a non-zero $\tenjsymbol^2$ (16162 and 308086 data points for the cutoffs 1.0 and 1.5 respectively). No transfer learning was applied. The figure below shows the training loss for both regressors.
\begin{figure}[h]
    \centering
    \includegraphics[width=0.65\linewidth]{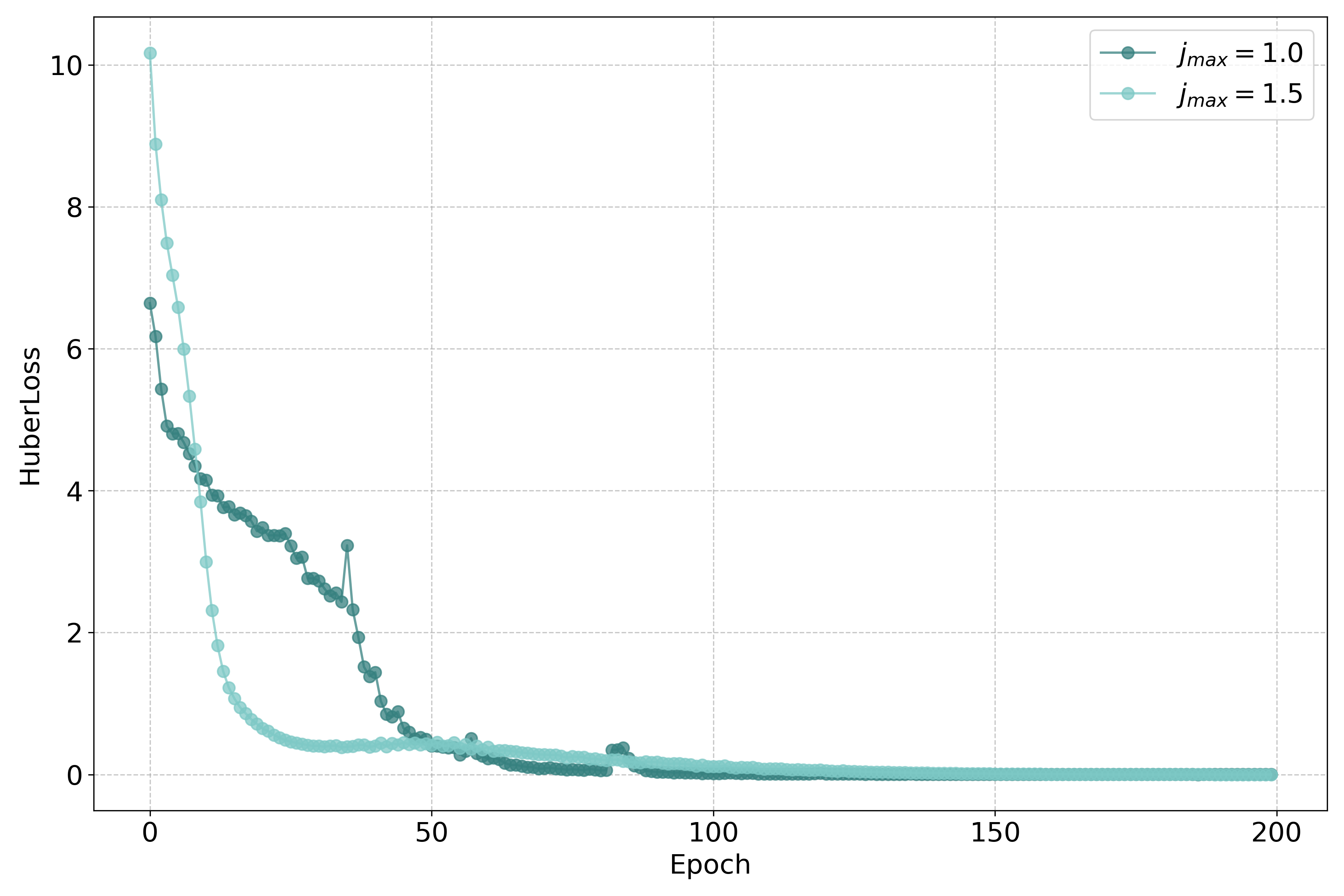}
    \caption{The training loss during the regressor training at cutoffs of 1.0 and 1.5}
    \label{fig:train_loss_reg}
\end{figure}
\\
As shown in Figure \ref{fig:train_loss_reg}, no irrecoverable or sustained spikes were observed. The regression metrics were observed to be as follows:
\begin{table}[h]
    \centering
    \begin{tabular}{cccccc}
        \rowcolor{palegreen!50}
        Training Cutoff & $\mathrm{MAPE}$ (\%) & $\mathrm{Acc}_{\leq 0.1}$ (\%) & $\mathrm{RMSE}_{\log}$ & $\mathrm{MAD}_{\log}$ & $R^2$ \\
        \hline
        1.0 & 2.7587 & 94.3045 & 0.0582 & 0.0215 & 0.9986 \\
        1.5 & 4.1735 & 83.6901 & 0.0851 & 0.0396 & 0.9999 \\
        \hline
    \end{tabular}
    \caption{Evaluation metrics on all non-zero configurations for the regressors trained at different cutoffs.}
    \label{tab:reg_cutoff_metrics}
\end{table}
\\
As shown in Table \ref{tab:reg_cutoff_metrics}, all metrics fall within a good range for such regression tasks. The regressor trained at a cutoff of 1.0 demonstrates excellent predictive accuracy, achieving a $\mathrm{MAPE}$ of 2.7587\% and a threshold accuracy $\mathrm{Acc}_{\leq 0.1}$ of 94.3045\%, indicating that the vast majority of predictions fall within a 10\% relative error margin of the true values when evaluating the model on all possible configurations which yield non-zero labels at the cutoff. Further, its $\mathrm{RMSE}_{\log}$ and $\mathrm{MAD}_{\log}$ values reflect a low dispersion of residuals and highlight the consistency of the regressor's output. The high $R^2$ score implies that nearly all variance in the target variable is captured by the regressor.
\newparagraph
In comparison, the model trained at $j_{max} = 1.5$ exhibited slightly less accurate performance, but nevertheless yield promising results. The $\mathrm{MAPE}$ and $\mathrm{Acc}_{\leq 0.1}$ values indicate that it maintains a good predictive fidelity. While the $\mathrm{RMSE}_{\log}$ and $\mathrm{MAD}_{\log}$ values are comparatively higher, they are still within acceptable bounds. Notably, the $R^2$ value for this regressor is even higher, indicating a better fit for the data. To further elucidate upon the performance of both regressors, a True vs. Prediction plot can be shown below.
\begin{figure}[h]
    \centering
    \begin{subfigure}[t]{0.495\textwidth}
        \centering
        \includegraphics[width=\textwidth]{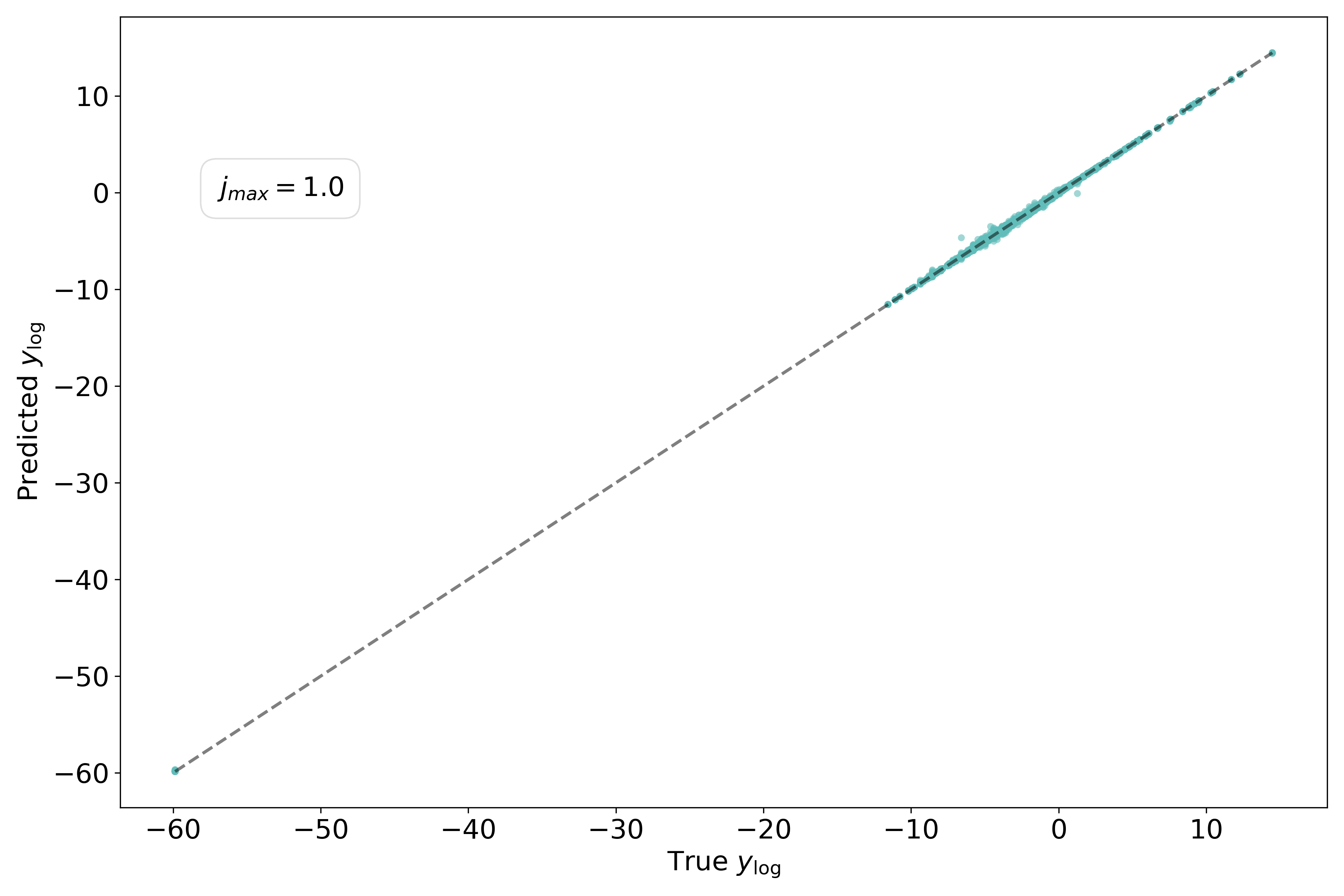}
        \caption{}
        \label{fig:subfig_FP}
    \end{subfigure}
    \hfill
    \begin{subfigure}[t]{0.495\textwidth}
        \centering
        \includegraphics[width=\textwidth]{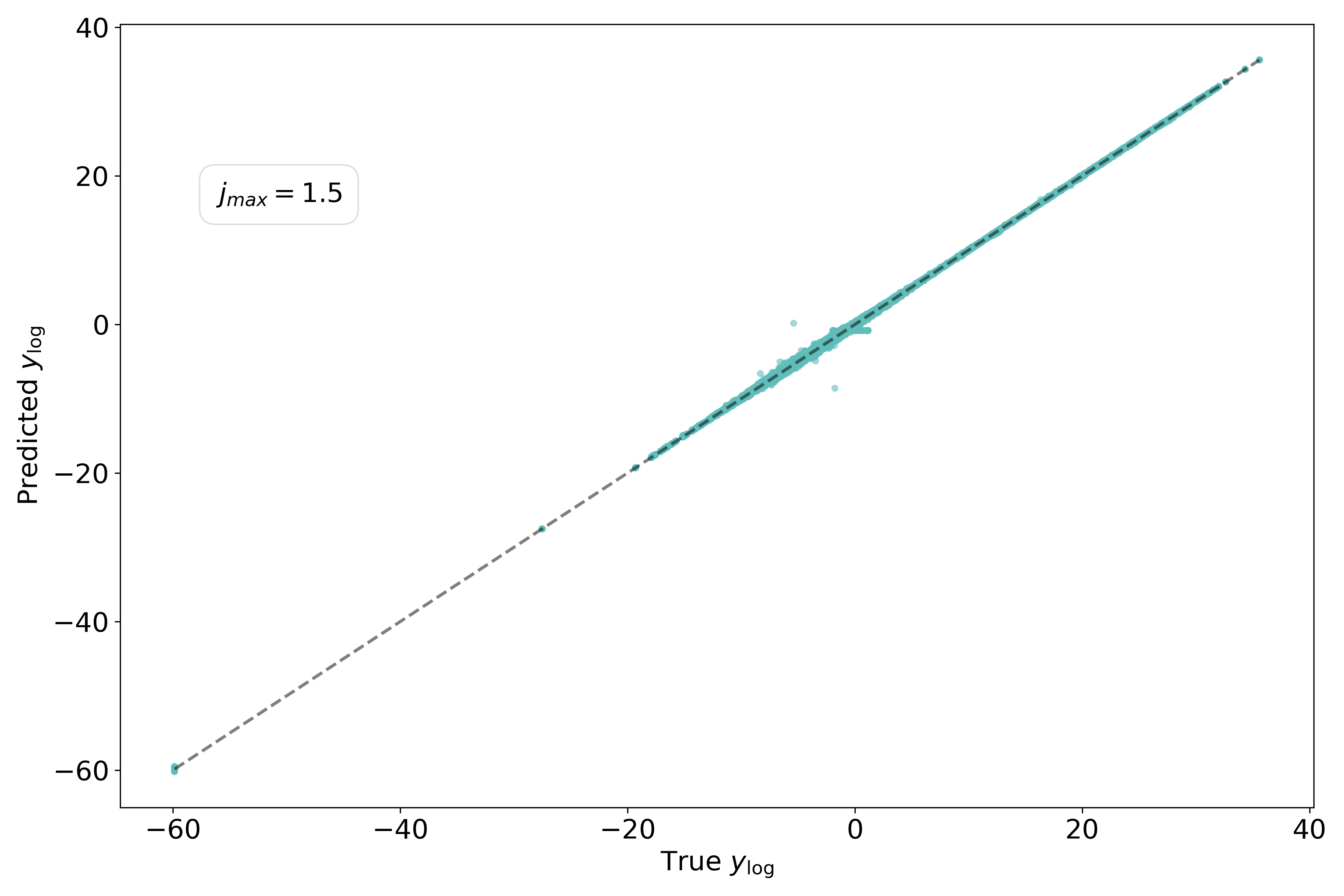}
        \caption{}
        \label{fig:subfig_FN}
    \end{subfigure}
    \caption{A True vs. Prediction plot in log space for the regressors trained at cutoffs 1.0 and 1.5 on the left and right respectively. The data points used constitute all data points at the respective cutoffs which yield non-zero labels.}
    \label{fig:TvP}
\end{figure}
\\
Figure \ref{fig:TvP} shows the True vs. Prediction plots in log space for both regressors at the cutoff 1.0 and 1.5 on the left and right respectively. The data points evaluated in the plots constitute all data points at the cutoff which yield a non-zero label. As shown, most of the predicted labels aligns well with the true labels in both cases, with only a few predictions which fall far from the true label values. Overall, the regressors seem to be performing relatively well. No minimal training dataset investigation or transfer learning was applied in both cases. We also note that the trained regressors did not generalise well beyond their training domain. This is to be expected, as extrapolation is generally a non-trivial task and is made more difficult by the nature of the vertex amplitude functions being highly oscillatory in general.

\subsubsection{Expert network}
So far, the task of computing the vertex amplitude using neural networks has been divided into a classification and regression tasks. Further, we have focused on learning the sign of the $(\tenjsymbol(\mathbf{S^{(j_{max})}}))^2$ for the classification task and the value of $\logtenj{\mathbf{S^{(j_{max})}}}$ for the regression task. To produce the correct amplitude value for a given spin configuration, we need to:
\begin{itemize}
    \item[(i)] at inference time, exponentiate the regressor's output to obtain $(\tenjsymbol(\mathbf{S^{(j_{max})}}))^2$ instead of $\logtenj{\mathbf{S^{(j_{max})}}}$ and then further take the square root to obtain the correct $\tenjsymbol(\mathbf{S^{(j_{max})}})$ value, 

    \item[(ii)] insert the correct sign factor based on the spins in the given $\mathbf{S^{(j_{max})}}$ as shown in equation \eqref{eq:10j_to_modified_10j} to obtain the correct sign of the computed $\tenjsymbol$, and lastly

    \item[(iii)]  insert the correct positive dimensionality multiplicative factor related to the spins in the given $\mathbf{S^{(j_{max})}}$ as shown in equation \eqref{eq:bc_va} to compute the complete vertex amplitude of the BC-model.
\end{itemize}
As such, the last piece of the work includes creating a meta network, denoted $\meta(\mathbf{S})$, which combines both the classifier and the regressor. The output of this meta network, which we shall call an Expert, is merely the product of the outputs of $\classifier(\mathbf{S})$ and $\regressor(\mathbf{S})$ along with all the relevant corrections mentioned above.
\newparagraph
The reason for this is not only aesthetic. The grouping of such networks into an Expert $\meta$ allows for having an ensemble of Experts for a given given cutoff, each trained with different seeds and therefore different datasets, in hope to increase the predictive accuracy of the overall model and produce more precise error estimation. Further, one can combine Experts in a Mixture of Experts (MoE) approach \cite{Jacobs:1991,Shazeer:2017} which, in simple terms, houses within it $M$ experts (or $M$ ensembles of $N$ experts each) each trained at a different cutoff. After appropriate training of the gating in such an MoE, this then results in one model which can accept any configuration which falls within the range of cutoffs it has been trained on and yield an accurate prediction based on the Experts it contains. This, however, will be left for future work.


\section{Discussion}
\label{sec:discussion}
One of the most, and perhaps the most, pressing inherent issue in this work is the scarcity of data. Generally, SL is a greedy approach and this only gets worse if the objective we attempt to learn is complicated (e.g. very large range, highly non-uniform, very sensitive to the inputs). This was already observed during the training of the regressor in this simple toy model. Different training methods, including transfer learning, different one-hot encoding and dataset processing, resulted either catastrophic forgetting or low predictive accuracy. This, however, can of course be due to the network architecture. Nevertheless, the issue of the regression task being data greedy is expected to still persist. In that case, one needs to tailor the networks such that they require as minimal data points as possible. Further, data collection can be conducted by Reinforcement Learning (RL) methods. During this work, we have also collected data points by training an agent with a proximal policy optimisation algorithm (PPO) \cite{Schulman:2017} to find the highest valued amplitudes without exhaustive enumeration of all possible configurations. This can then cut down on unnecessarily computing vertex amplitudes which may be irrelevant to the desired training process, albeit still being a computationally expensive process. 
\newparagraph
Spinfoam vertex amplitudes, for a given set of coherent states as boundary data, have been computed using Monte Carlo methods \cite{Steinhaus:2024qov}. One may be able to adapt the generative flow networks approach \cite{Bunao:2024qwm} (initially used to compute expectation values of observables by learning the regions in which the amplitude is large) to facilitate a similar computation as done in \cite{Steinhaus:2024qov}. Such flow networks may also provide another ``agent"-like approach for data collection in the domain of direct learning of the vertex amplitude itself, as we have carried out. How to exactly set it up is not clear at the current time, but an interesting avenue to explore in future work.
\newparagraph
Ultimately, we recognise that this work serves only as a proof-of-principle, and thus we refrained from exploring or utilising all possible avenues to identify, resolve or optimise such issues and bottlenecks. This is left to be conducted for later work on more physically relevant models but we are aware that this is a persistent issue in the nature of this approach. The resolution of this issue will also largely rely in part to utilising other efficient numerical methods to generate sufficient amount of data for training as this sets the bound of the amount of data available for training.
\newparagraph
The next pressing issue is the regression problem. The learned data for the regressor in this work spanned a very large range which only grew with the cutoff. This is due to the nature of the learned amplitude as they are generally represented by highly oscillatory functions. This poses several serious concerns. For example, assuming the data acquisition process is not an issue (may already be not true for more realistic models such as the EPRL model at high spin), devising a training set which has enough representatives from each order of magnitude is a non-trivial task. This will also highly depend on the network and loss function used as different architectures and losses can be less sensitive to very large data while others might require more large data representatives in the dataset. It is therefore easy to see how this becomes a concern if one can not enumerate the entire space. 
\newparagraph
In the cutoffs presented in this work, the classifier maintained a static architecture which included a number of trainable parameters $\texttt{Params}(\classifier)$ of only 1537 parameters across all training. Despite that, it was demonstrated that it excelled in learning whether the given spin configuration would yield a non-zero squared $\tenjsymbol$ or not. The regressors on the other hand had a number of trainable parameters $\texttt{Params}(\regressor)$ of 350093 and 1091725 for $j_{max} = 1.0$ and $1.5$ respectively. Compared to the total number of non-zero data points available at the same cutoffs (19015 and 362455 respectively), one sees that this approach is simply inefficient: with enough learnable parameters, one can fit any data. In this case, it is much faster to simply create a table of all possible amplitude values for all spins in the current cutoff. If a regressor trained on some cutoff $j_i$ can, to some degree of acceptable evaluation metrics, predict on a higher cutoff $j_k > j_i$, then $\texttt{Params}(\regressor)$ being larger than the number of the available states at the training cutoff can be overlooked. This, however, is not the case in this work. 
\newparagraph
Nevertheless, that does not mean that it is not possible. Ablation studies (here discussed for $\jmax = 1.0$) were conducted where different MLPs with different widths and depths were tested. It was observed that one can get to moderately acceptable evaluation metrics with MLPs of depth 3 and width 64 ($\texttt{Params}(\regressor) = 10685$) and even MLPs with depth 1 and width 256 ($\texttt{Params}(\regressor) = 8253$). While in both cases $\texttt{Params}(\regressor)$ is less than the total number of available data points, this does not immediately translate to all relevant evaluation metrics being consistently high or the training process being conducted smoothly. Further, different architectures, such as recurrent neural networks with gating\footnote{networks with gated recurrent units (GRUcells) \cite{Cho:2014}} were observed to be good candidates for the task. Lastly, given the graph based nature of the 10$j$ symbol, graph neural networks (GNNs) \cite{Scarselli:2009} might also serve as a good candidate. However, this work has not explored such an architecture. Lastly, the regressors in this work did not produce satisfactory evaluation metrics when evaluated beyond the training domain. This is unsurprising, as this is an inherent limitation to such tasks which is further made difficult by the objective function being learned, here the vertex amplitude, being of difficult nature. 
\newparagraph
This leads to the following conclusion: while the regressors used in this work are inefficient, \emph{in principle}, we believe that there are more efficient architectures to be explored with $\texttt{Params}(\regressor)$ being less than the available training data. What they are, how well they train and whether they, if at all possible, generalise for higher cutoffs or not were not tasks of principal importance in this work. The reason being that it is unclear precisely how the tools developed in this work would translate to physically relevant models such as the EPRL model. The purpose of this work is to merely demonstrate a proof-of-principle and exhaustive studies will be left for later work.

\section{Conclusion}
Spinfoam theories provide dynamics for non-perturbative loop quantum gravity by constructing transition amplitudes between spin-network states through a sum over their histories. The quantisation procedure implements simplicity constraints at the quantum level, resulting in a regularised partition function $Z_{\Delta^*}$ on a spinfoam $\Delta^*$. One of the main components of $Z_{\Delta^*}$ is the vertex amplitude $A_v$ which, akin to QED, encodes the local dynamics of quantum geometry. In this work, we investigated the feasibility, as a proof-of-principle, of a data-driven approach whereby the vertex amplitude $A_v^{BC}$ of the Euclidean Barrett-Crane model is learned using deep neural networks, specifically the Riemannian 10$j$ symbol which is at the core of $A_v^{BC}$ in this model. The amplitude is learned through a two-step process whereby first, a classifier $\classifier$ is trained to predict whether for a given set of spin configurations $\mathbf{S}$, the resulting $A_v^{BC}(\mathbf{S})$ is zero or not. Second, a regressor $\regressor$ is trained to predict the exact numerical value of the $\tenjsymbol^2$. As a last step, we construct a meta network, which we denote an Expert $\mathcal{P}$, which utilises both $\classifier$ and $\regressor$ to construct the correct amplitude $A_v^{BC}$ by inserting the relevant sign and dimensionality factors.
\newparagraph
For the classification task, a small MLP was trained on several cutoffs ranging from 0.5 to 2.0, each time utilising transfer learning. Despite the relative training dataset size being reduced from 75\% to roughly 1\% of all available states for cutoffs of 0.5 to 2.0 respectively, the classifier $\classifier$ proved successful by being able to have high evaluation metrics (soft accuracy, hard accuracy, precision, recall and F-1 score) both within and well above the training cutoffs. The regression task proceeded with MLPs of dynamic architecture which depended on the training cutoff. Within the learned cutoff, the regressors showed high evaluation metrics (RMSE, MAD in log space, MAPE and threshold accuracy with $\epsilon = 0.1$) in both cutoffs presented (1.0 and 1.5). Generalisation to cutoffs beyond the training regime for the regressor case, however, was unsuccessful. 
\newparagraph
An Expert $\mathcal{P}$ was constructed to correctly output the full amplitude $A_v^{BC}$. We also discuss the limitations and hurdles encountered during this work, mainly the generalisation of the regressor to higher cutoffs and the training process with limited data. We propose an ensemble approach to increase the accuracy for the trained cutoffs and a reinforcement learning inspired approach to collect relevant training data where an agent is trained using a proximal policy optimisation algorithm to learn which configurations $\mathbf{S}$ would yield amplitudes relevant to the training process at hand. We concluded this work by elucidating upon different network architectures which may be better suited for the regression task. However, as this work stands as a proof-of-principle, we refrain from exploring all avenues to resolve the issues encountered in this work, this will be pursued later for physically relevant models. Nevertheless, the current work stands as an addition to existing numerical methods to compute the vertex amplitude of spinfoam models and provides a proof-of-principle that vertex amplitudes of spinfoam models are amenable to modern deep learning techniques. The aim of the surrogate models to be developed in this data-driven approach is to complement the numerical implementations of exact analytical methods by helping identifying dominant configurations, guiding importance sampling and Monte Carlo approximations and enabling efficient pre-selection in possibly large parameter spaces.

\ack
HS thanks Wojciech Kaminski and Martin Zei{\ss} for many helpful conversations about spinfoam models. HS acknowledges the contribution of the COST Action CA23130. The authors gratefully acknowledge the scientific support and HPC resources provided by the Erlangen National High Performance Computing Center (NHR@FAU) of the Friedrich-Alexander-Universität Erlangen-Nürnberg (FAU). The hardware is funded by the German Research Foundation (DFG).


\end{document}